\documentclass[a4paper, 11pt]{article}
\usepackage{amsmath, amsfonts, amsthm, mathrsfs, graphicx, cite}

\title{A Formal Transaction Cost-Based Analysis of the Economic Feasibility of Ecosystems}
 
\author{Christoph F. Strnadl\footnote{Software AG, CTO Office, Vienna, Austria; \texttt{christoph.strnadl@softwareag.com}; ORCID: 0000-0003-4173-656X}}

\date{January 28th, 2024}

\begin{document}

\maketitle

\begin{abstract}

Ecosystems enjoy increasing attention due to their flexibility and innovative power. It is well known, however, that this type of network-based economic governance structures occupies a potentially unstable position between the two stable (governance) endpoints, namely the firm (i.e., hierarchical governance) and the (open) market (i.e., coordination through the monetary system). 

This paper develops a formal (mathematical) theory of the economic value of (generic) ecosystem by extending transaction costs economics using certain elements from service-dominant logic. 

Within a first-best setting of rational actors, we derive analytical solutions for the hub-and-spoke and generic ecosystem configurations under some uniformity assumptions of ecosystem participants. Additionally, we are able to infer a generic condition for the welfare-maximizing and utility-maximizing price of the hub-and-spoke configuration in the familiar form of Lerner indices and elasticities.

Relinquishing a first-best rational actors approach, we additionally derive several general propositions on (i) necessary conditions for the economic feasibility of ecosystem-based transactions, (ii) scaling requirements for ecosystem stability, and (iii) a generic feasibility condition for arbitrary provider-consumer ecosystems. 

Finally, we present an algebraic definition of business ecosystems and relate it to existing informal definition attempts. Thereby we demonstrate that the property of ``being an ecosystem'' of a network of transacting actors cannot be decided on structural grounds alone.

\end{abstract}

\newline


\pagebreak

\tableofcontents

\section{Introduction}

Economic activities are typically organized either in firms, i.e., using a hierarchical form of governance, or as markets using (product and service) prices as coordination mechanism. Transaction cost economics (TCE) \cite{Williamson-1985} explains that very well by reformulating the question of the optimal organizational form for a particular set of economic activities against the backdrop of expected transaction costs. TCE has further developed criteria and conditions indicating the respective preferences for either hierarchical or market-driven organization such as asset specificity, frequency of transactions, and others. 

Recently, though, intermediate forms of governance between the polar opposites of firm and market such as network systems \cite{Beck-2006} and ecosystems  \cite{Hein-2020} have increasingly garnered attention. Due to the looser coupling of actors compared to a single firm or a tightly integrated vertical supply chain, ecosystems promise to solve pressing business problems in today's highly competitive and global markets related to the following activities:

\begin{itemize}
\item creating complex product or service offerings consisting of several (complementary) elements
\item innovation under rapidly changing conditions where the innovating firm does not possess all required capabilities and skills
\item reduce distribution costs and complexity by creating a novel, more efficient channel
\item disrupt an existing industry by reaping economies of scale and scope by partnering with existing players
\end{itemize}

Companies with 50\% or more revenues from ecosystems with concomitant better understanding of customer needs or ecosystem-driving companies then are reported to enjoy a 20-27\% higher profit margin and 27-32\% revenue growth increase compared to industry average \cite{Weill-2015, Fuller-2019}.

For our purposes we will loosely make use of the ecosystem definitions of Adner \cite{Adner-2017} and Jacobides \cite{Jacobides-2018}:

\begin{quote}
[An ecosystem is] the alignment structure of the multilateral set of partners that need to interact in order for a focal value proposition to materialize. \cite{Adner-2017}
\end{quote}
\begin{quote}
An ecosystem is a set of actors with varying degrees of multi\-la\-teral, non-generic complementarities that are not fully hierarchically controlled. \cite{Jacobides-2018}
\end{quote}

The sharing of data across organizations has recently received increasing attention. While established concepts such as databases, data warehouses (on premise and in the cloud), data lakes, data lakehouses and, latest, data meshes primarily address the exchange of (mainly analytical) data within organizations, the focus is now shifting to so-called data platforms for the exchange of operational and analytical data across organizations under terms such as data space and (data) ecosystem.

The EU has designated the free flow of data as the next (fifth) freedom of the Union expecting that the ensuing data economy will significantly contribute to the EU's economic success \cite{ECom-2020}. In this vein, data platforms and data spaces present a new (promising) way to create ecosystems based on frictionless (i.e., entailing very low transaction costs) data exchange and, thus, enabling new, data-based business models. For our purposes, a data space may be defined as follows:
\begin{quote}
A data space is a coordinated set of technical standards, organizational policies, and core services under a specified governance model to enable and facilitate data exchange between its par\-ti\-ci\-pants.\cite{Strnadl-2023a} 
\end{quote}
Some authors and institutions, notably the European Parliament and the European Council in their recent Data Act proposal \cite{ECouncil-2023}, also include joint data processing in their definitions of data spaces. This extends well into the realm of data \emph{ecosystems}. To the extent that the value proposition is more generally based on the exchange of (IT-mediated) services instead of being confined to the exchange of data, one may speak of a service ecosystem such as Gaia-X \cite{Gaia-X-2022a, Gaia-X-2022b}.

Our contribution develops the theory of economic feasibility of arbitrary ecosystems in an inductive fashion in section \ref{sec-theory}. Only at the end of the section will we present the generic algebraic definition of a business ecosystem and the \emph{master equation} governing its economic feasibility. Section \ref{sec-discussion} discusses certain assumptions and implications of our approach. There, we finally show how our formalization complements current informal definition attempts of business ecosystems.

\section{Theory}			\label{sec-theory}
\subsection{Transaction Feasibility}

\newtheorem{definition}{Definition}
\newtheorem{proposition}{Proposition}	
\newtheorem{lemma}{Lemma}
\newtheorem{corollary}{Corollary}

\newcommand{\mE}{\mathscr{E}}

Let $T = I + E $ denote the transaction costs for one participant of a transaction\footnote{We will assume only 2-party transactions here.} \cite{Williamson-1985}.
Here, $I$ denotes the \emph{ex ante} transaction costs such as the costs for searching, bargaining, or contracting, and $E$ represents \emph{ex post} transaction costs such as policing and enforcement costs, but also opportunity costs in cases where the transaction does not meet expectations \cite{Dahlman-1979, Chang-2000}. Let sub- or superscripts $C$ and $P$ denote data or services \emph{consumers} and \emph{providers}, respectively, and $X$
the (usually monetary) value exchanged during the data sharing or service consumption interaction, i.e., the money the consumer pays to the provider for the data or use of service.

At this point we extend a typical TCE-based setting and introduce an additional variable $V$ for both providers and consumers denoting the expected gross total internal value or benefit of a successful interaction. This step is motivated by employing service-dominant logic \cite{Vargo-2020} which explicitly recognizes the co-creation of value for both partners during a service-type interaction. A similar step has been recently taken by \cite{Meierhofer-2022, Meierhofer-2023}, although they limit the application to single binary provider-client interactions and determine the Pareto-efficient optima in the two-dimensional space spanned by $V_P$ and $V_C$. As we shall see later, the level of $V$ will critically influence ecosystem feasibility acting as a kind of cohesive force for the ecosystem.

Finally, $W$ denotes the expected total net internal value of the transaction for the respective participant, i.e., the expected gross value less transaction costs $T$ and less transaction fee for the consumer or including the transaction fee for the provider. 

Then, in the absence of exogenous incentives\footnote{such as governmental subsidies or mandatory regulatory requirements}, the following two \emph{master equations} define the economic feasibility for the transaction to take place\footnote{In the principal-agent model, similar conditions are called participation conditions \cite{Laffont-2002}.}.
\begin{eqnarray}
	\mathrm{Provider:} \; & W_P = V_P + X - T_P & > \ 0, 	\label{eqn-provider-feasibility}		\\
	\mathrm{Consumer:} \; & W_C = V_C - X - T_C & > \ 0.	\label{eqn-consumer-feasibility}
\end{eqnarray}

Because $X$ and $T$ are typically specified in monetary units whereas $V$ is often given relative to a utility scale private to each participant, we assume the existence of an inter-personal common utility scale for all participants and that all variables have been suitably transformed to this scale in our formulas. We are well aware of the fact that this has been a highly contentious assumption since the beginnings of game and decision theory (see, e.g., already \cite{Luce-1957}) and continues to be so \cite{Kaneko-1984, Harsanyi-1991, Fleurbaey-2004}. However, at this point in the development of the theory we believe this to acceptable.

We note that these equations only present \emph{necessary} conditions for the exchange to take place, but are not sufficient. 

In data ecosystems, the terms \emph{consumer} and \emph{provider} also denote the logical flow of data which gets transferred from the data provider to the data consumer or is made accessible to the data consumer irrespective of how this is technically achieved\footnote{It may very well be that the data provider uses a facility provided by the data consumer to "upload" their data to the consumer.}. In the case of service ecosystems, consumers \emph{consume} the services offered by providers.

In the case of a pure data exchange, the additional value $V_P$ of such a transaction for a data provider cannot lie in the shared data alone as this data is already in the possession of the data provider before and without any transaction taking place. Hence, any $V_P > 0$ needs to come from other (fringe or later) benefits of the transaction. The same holds for service providers which, in general, do not profit from the service they sell to other consumers beyond the service fee they receive. We refer the reader to section \ref{sec-discussion} for a discussion and examples.

By adding eqn.s (\ref{eqn-provider-feasibility}) and (\ref{eqn-consumer-feasibility}) we obtain the first overall feasibility condition for an ecosystem transaction to happen.
\begin{equation}
	V_P + V_C - (T_P + T_C) > 0.		\label{eqn-tx-feasibility-general}
\end{equation}

This means that the total (gross) value generated by the transaction for both provider and consumer must (at least) cover both their transaction costs. 

Assume now that there is no further value of the data sharing transaction for the data provider, that is, $V_P = 0$. Then we can immediately derive the following \emph{subsidization theorem}.

\begin{proposition}[Subsidization]	\label{th-subsidization}
In the absence of \emph{ex ante} expected benefits (i.e., $V_P = 0$) in excess of the explicit data exchange fee $X$, the data consumer has to fully subsidize all transaction costs of the data provider to render the transaction economically feasible.
\begin{equation}
	V_C > T_P.
\end{equation}
\end{proposition}

\begin{proof}
Setting $V_P = 0$ in eqn.~(\ref{eqn-provider-feasibility}) and observing that $T_C > 0$ it follows that
\begin{equation*}
	V_C - (T_P + T_C ) > 0	\quad \Rightarrow \quad V_C > T_P + T_C > 0 \quad \Rightarrow \quad V_C > T_P.
\end{equation*}
\end{proof}

Note that this result is valid for any general data transaction irrespective of whether such data exchange is conducted within a data space or data ecosystem or not (e.g., using ordinary point-to-point file transfer mechanisms). Because eqn.~(\ref{eqn-provider-feasibility}) reduces to $X > T_P$ in this case, it also immediately follows that
\begin{equation}
			V_C > X > T_P.
\end{equation}
That implies that data consumer benefits $V_C$ must fund a transaction fee $X$ covering at least the transaction costs of the data provider.

\subsection{Analytic solution of the two-actor case}		\label{seq-analytic-sol-2-actors}

We will now derive an analytic solution of the two actor case as described by eqn.s (\ref{eqn-provider-feasibility}) and (\ref{eqn-consumer-feasibility}) under the assumptions of rational, i.e., utility maximizing, economic actors for the first best case, that is, when all actors have full information about the other actors decision context and strategy. As we will be able to show later, proposition \ref{p-eco-sol-2-actors} of this section will play a central role for the analytic solution of the other configurations.

The following definition captures the assumptions on \emph{rational ecosystem actors} (as we call them):

\begin{definition}[Rational ecosystem actors]		\label{def-rat-eco-actors}
Let $P$ and $C$ be economic (ecosystem) actors with $W^P$ and $W^C$ the utility (value) functions for the re\-spec\-tive actors for carrying out a mutual transaction. Then we call both, $P$ and $C$, \emph{rational ecosystem actors} iff they mutually transact according to the foll\-owing conditions:

\begin{enumerate}
\item $P$ and $C$ transact iff $W^C > 0$ and $W^P > 0$.			\label{def-rat-eco-actors-1}
\item Neither $P$ nor $C$ have alternatives, complementarities, or substitutes for conducting the transaction.
\item $W^P$ and $W^C$ fully capture the preferences of both actors. That means, there are no hidden motives like in the ultimatum game, in cases of revenge or altruistic behavior.
\item Both actors act according to the principle of maximizing their utility (here: $W^P$ and $W^C$) from the trade.
\end{enumerate}
\end{definition}

Before presenting the solution we observe that in the master equations above the transaction fee $X$ is the only economic variable left to be decided before the trade or exchange can take place. Then the following proposition gives an analytic solution to this decision problem under certain conditions and assumptions.

\begin{proposition}[2-actor ecosystem solution]		\label{p-eco-sol-2-actors}
Let $P$ and $C$ be two rational ecosystem actors as per definition \ref{def-rat-eco-actors}, and let the following two (master) equations govern the conditions under which a mutual transaction may take place.
\begin{eqnarray}
	\mathrm{Provider:~} P \; & W^P = V^P + X - T^P & > \ 0, 	\label{eq-eco-sol-th-prov}		\\
	\mathrm{Consumer:~} C \; & W^C = V^C - X - T^C & > \ 0.		\label{eq-eco-sol-th-cons}
\end{eqnarray}
Then, in a first best scenario, the transaction takes place under the condition
\begin{equation}							
	V^C + V^P  >  T^C + T^P										\label{eq-eco-sol-cond}
\end{equation}
and the transaction fee $X^\star$ is given by
\begin{equation}												\label{eq-eco-sol-x}
X^\star = \frac{1}{2} \left[ (V^C - V^P) - (T^C - T^P) \right].		
\end{equation}
\end{proposition}
\begin{proof}
By assumption, both $P$ and $C$ try to maximize their value function:
\begin{eqnarray*}
		\max W^P = V^P + X - T^P, 			\\
		\max W^C = V^C - X - T^C.	
\end{eqnarray*}
subject to the two conditions of the master equations. In recognition that the only free economic variable is the transaction fee $X$, this problem is equivalent to $\max X$ for the provider and $\min X$ for the consumer. Now, under first best conditions, both actors know the other actor's utility function (here $W^P$ and $W^C$) and strategy context. This may be summarized as the following optimization problem:
\begin{equation}
\begin{array}{rl}
	\mathrm{Provider:}& 	\; \max X,				\label{eq-sol-opt-problem}\\
	\mathrm{Consumer:}& \; \min X,		
\end{array}
\end{equation}
subject to
\begin{equation}							\label{eq-sol-space-for-x}
	T^P - V^P < X < V^C - T^C.
\end{equation}

Now, under first best conditions both players know the others utility function and strategy, as given in equ.s (\ref{eq-sol-opt-problem})-(\ref{eq-sol-space-for-x}). In particular, they know that the admissible bargaining range for $X$ is given by eqn.~(\ref{eq-sol-space-for-x}). They also know their competing strategies regarding $X$. Hence, in absence of any alternatives to making the trade with another partner (as per the rational ecosystem actors assumption), and in view of the symmetric situation, both rational players will agree to ``meet in the middle'' regarding the transaction fee $X^\star$.

That means we can write $X^\star$ as follows, proving eqn.~(\ref{eq-eco-sol-x}):
\begin{eqnarray*}
	X^\star & = & T^P - V^P + \frac{1}{2} \left[ V^C - T^C - (T^P - V^P) \right]		\\
	        & = & \frac{1}{2} \left[ (V^C - V^P) - (T^C - T^P) \right].
\end{eqnarray*}
Inserting $X^\star$ from eqn.~(\ref{eq-eco-sol-x}) into the two equations $W^P > 0$ and $W^P > 0$ as stated in condition \ref{def-rat-eco-actors-1} of definition \ref{def-rat-eco-actors} directly yields eqn.~(\ref{eq-eco-sol-cond}).
\end{proof}

We will use this 2-actor solution later to derive analytical expressions for the hub and generic ecosystem configurations.

It is also interesting to note that $X^\star < 0$ is possible under the condition that the difference of consumer and provider values does not recover the total transaction cost.

\subsection{Ecosystem Preference Analysis}

Let us now turn to a situation where participants can choose whether the (data or service) transaction takes place outside a data space or ecosystem or within such an ecosystem, a situation typical when more than one technology fOEM
solution for executing the transaction is available to both participants. For instance, in the case of Gaia-X, participants have the option of doing data exchange based on classical (point-to-point) file transfer protocols or using Gaia-X mechanisms and standards.
We will use the superscript `$\star$' to denote variables pertaining to an ecosystem-based transaction (e.g., Gaia-X) and no superscript for the other type of transaction mechanics. 
Assuming that the internal value $V_C$ of the transaction for the consumer does not depend on the actual exchange mechanism and, hence, the transaction fee $X$ does not change either, the two financial feasibility equations (\ref{eqn-provider-feasibility}) and (\ref{eqn-consumer-feasibility}) for an ecosystem-based transaction read as follows.
\begin{eqnarray}
	\mathrm{Provider:} 	& W_P^\star = V_P^\star + X - T_P^\star & > 0, 	\label{eqn-provider-feasibility-eco}		\\
	\mathrm{Consumer:}	& W_C^\star = V_C - X - T_C^\star & > 0.		\label{eqn-consumer-feasibility-eco}
\end{eqnarray}

Participants will only conduct a transaction via the ecosystem if its respective net value $W^\star$ is greater than if conducted outside an ecosystem, i.e., if $W^\star > W$. This quickly leads to two propositions defining necessary conditions on consumer and provider engagement.

\begin{proposition}[Data consumer engagement]\label{th-data-cons-engage}
A data consumer $C$ will prefer ecosystem-based over standard transactions only if ecosystem-based transaction costs are lower than those for standard transactions, that is, $T_C^\star < T_C$.
\end{proposition}
\begin{proof}
Condition $W_C^\star > W_C$ directly translates into
\begin{equation}
	V_C + X - T_C^\star > V_C + X - T_C \ \Rightarrow \ - T_C^\star > - T_C \ \Rightarrow \  T_C^\star < T_C.	\label{eqn-eco-scale-consumer}
\end{equation}
\end{proof}
This condition (\ref{eqn-eco-scale-consumer}) may be achieved by suitable economies of scale of an ecosystem, for instance.

The situation for data providers is slightly different because the intrinsic value of the data exchange may be different for ecosystem-based and standard forms of data exchange as we will show by analyzing the following proposition.

\begin{proposition}[Data provider engagement]
A data provider $P$ will only prefer ecosystem-based over standard data exchange if the incremental value exceeds the incremental transaction costs.
\begin{equation}
	\Delta V_P^\star := V_P^\star - V_P > T_P^\star - T_P.		\label{eqn-eco-scale-provider}
\end{equation}
\end{proposition}
\begin{proof}
The proof is conducted similar to proposition \ref{th-data-cons-engage} using $W_P^\star > W_P$.
\end{proof}

Analyzing this, we discriminate four cases:

\begin{center}
\begin{tabular}[c]{ccc}
\hline
No. 	&	$V_P^\star - V_P$	&	$T_P^\star - T_P$	\\
\hline
(a)		&	$>0$				& 	$<0$	\\
(b)		&	$>0$				& 	$>0$	\\
(c)		&	$\le 0$				& 	$<0$	\\
(d)		&	$\le 0$				& 	$>0$	\\
\hline
\end{tabular} \\
\end{center}

Cases (a) and (c) are not very illuminating because condition $T_P^\star < T_P$, e.g., brought about by economies of scale or simply better technology,
will trivially entice providers to prefer ecosystem technology.

Case (d) also results in an obvious preference of the provider to not choose the ecosystem transaction mechanisms: Not only are transaction costs higher, $T_P^\star > T_P$, but the (expected internal) value $V_P^\star$ is also less than what they can expect using the standard transaction mechanism as $V_P^\star < V_P$.

This leaves case (b) as the most interesting situation where higher ecosystem transaction costs $T_P^\star$ are recouped by a higher level of expected internal benefits, $V_P^\star$.

\subsection{Ecosystem Feasibility - Hub Configuration}			\label{subsec-eco-via-hub}

\subsubsection{Initial Considerations}
Assume an ecosystem consisting of a single consumer $C$ and $n$ providers $P_1,...,P_n$ as might be found in supply chains where a strong OEM\footnote{original equipment manufacturer} is collecting product and logistics data from its many suppliers. In the absence of outside beneficial arrangements, providers $P_i$ will demand a transaction fee $X_i$ for this exchange to happen, whereas the consumer $C$ will expect an internal value of $V^C_i$ at transaction costs $T^C_i$ when interacting with provider $P_i$. As we are solely dealing with transaction \emph{within} an ecosystem, we drop the superscript `$\star$' for entities $V$ and $T$, in favour of superscripts $C$ and $P$ denoting that the variable pertains to a consumer or a provider, respectively
We also assume that data providers cannot expect any fringe benefit from their data sharing activities by setting $V^P_i = 0$. 
Let's also explicitly consider the \emph{ex ante} investment $I_C$ of the consumer.

\begin{quote}
{\small
The reader needs to be aware that in a stable ecosystem these variables do not pertain to a \emph{single} transaction 
but represent the aggregate values of several transactions between two partners over a certain period of time (which is the same for all ecosystem participants). Investment $I^C$ of the consumer then takes place before the very first transaction is being conducted. }
\end{quote}

In this case, the $n+1$ master equations are given by 
\begin{eqnarray}
	X_i - T^P_i & > & 0, \qquad i = 1,...,n											\label{eqn-provider-i}	\\
	\sum_{i=1}^n V^C_i - \sum_{i=1}^n X_i - \sum_{i=1}^n T^C_i - I^C &  > & 0. 		\label{eqn-consumer-i}
\end{eqnarray}

Summing up the $n$ replicas of eqn.~(\ref{eqn-provider-i}) and eqn.~(\ref{eqn-consumer-i}), we obtain the overall ecosystem feasibility equation

\begin{equation}
	\sum_{i=1}^n ( V_i - T_i ) - \sum_{i=1}^n T_i - I^C \ > \ 0.					\label{eqn-eco-platform}
\end{equation}

For our further analysis we will replace the individual terms in both sums in eqn.~(\ref{eqn-eco-platform}) by their (arithmetic) averages

\begin{equation}															\label{eqn-consumer-hub-n}
	\sum_{i=1}^n ( V^C_i - T^C_i ) = n \cdot ( \bar{V}^C - \bar{T}^C ),\qquad \sum_{i=1}^n T^P_i	= n \cdot \bar{T}^P,
\end{equation}
yielding the following feasibility condition linear in $n$, the number of pro\-vi\-ders:
\begin{equation}
	n \cdot ( \bar{V}^C - \bar{T}^C ) - n \cdot \bar{T}^P - I^C > 0.		\label{eqn-consumer-hub-via-n}
\end{equation}

\begin{quotation}
{\small
Note that, \emph{sensu stricto}, such a substitution is only correct at one particular value of $n$. In other words, in all generality, the averages 
$\bar{V}^C, \bar{T}^C$ and $\bar{T}^P$ would depend on $n$. We argue, though, that with increasing experience of data space or ecosystem participants, these average values will either converge or not fluctuate in such a way as to render our derivations void.
}
\end{quotation}

\begin{figure}[!tbp]
\begin{center}
\includegraphics[width=\textwidth]{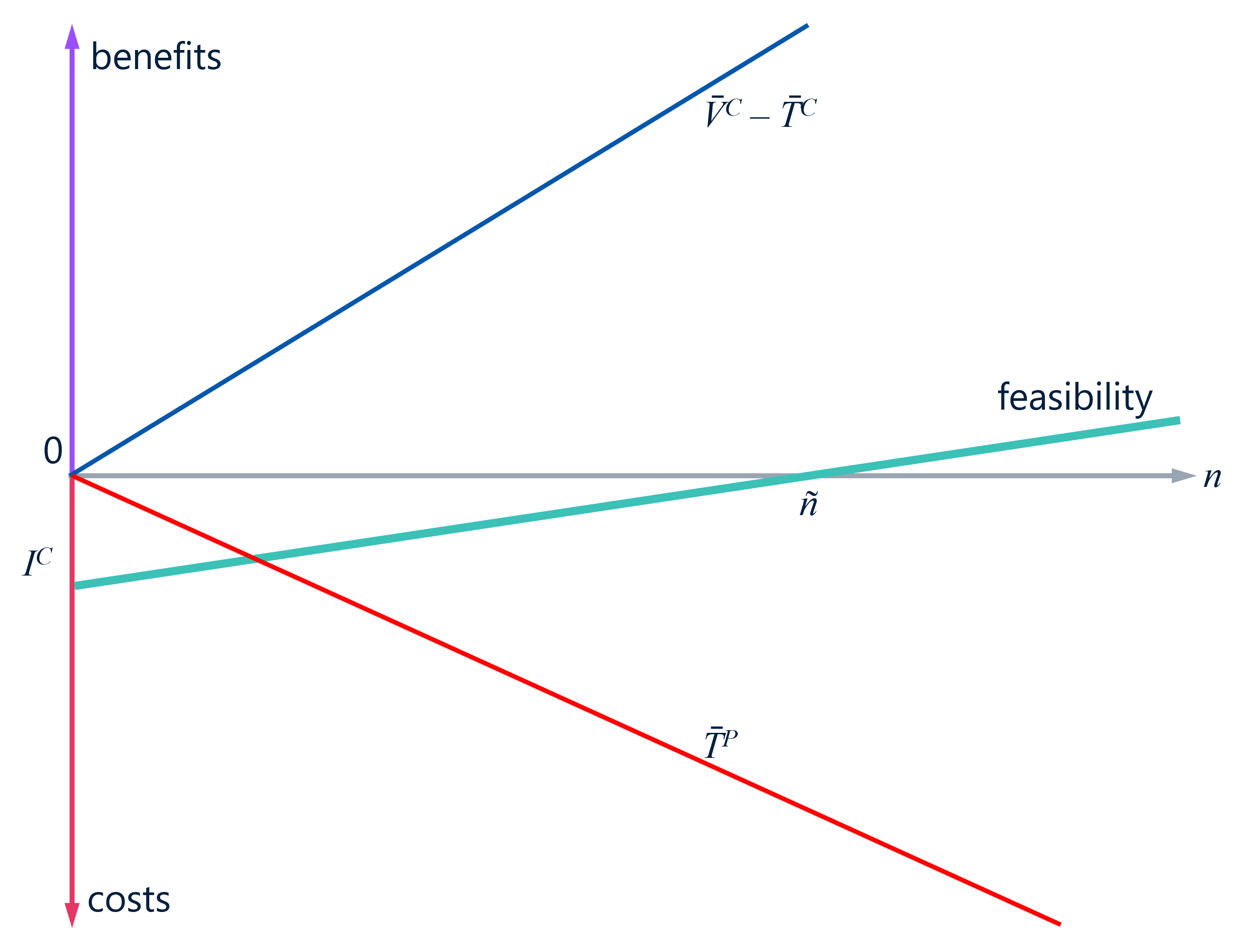}
\end{center}
\caption{Benefits/cost diagram for a hub configuration}
\label{fig-data-consumer-hub-cost}
\end{figure}

Furthermore, we assume that average net benefits for the consumer are always greater than average costs for providers, i.e., 
$ \bar{V}^C - \bar{T}^C  > \bar{T}^P$. Then figure \ref{fig-data-consumer-hub-cost} shows the overall feasibility of this ecosystem for different values of $n$. Because of the consumer's initial investment, $I^C$, this ecosystem only becomes (economically) feasible in case the number of data providers, $n$, exceeds a certain threshold value, i.e., $n > \tilde{n} = I^C / (\bar{V}^C - \bar{T}^C - \bar{T}^P)$.

\begin{quotation}
{\small
One might wonder whether the linear relation in the number of participants in eqn.~(\ref{eqn-consumer-hub-n}) may be relaxed by a more generic expression in form of two functions $\alpha(n) = \sum_{i=1}^n (V^C_i - T^C_i)$ and $\beta(n) = \sum_{i=1}^n T^P_i$ with $\alpha'(n) > \beta'(n)$. This is tantamount to replacing the condition relating average consumer benefits and provider costs with incremental benefits and costs.

Sadly, though, that is not sufficient because one cannot guarantee the existence of an $n$ such that $\int_0^n \alpha'(x) - \beta'(x) \; \mathrm{d}x > I_C$. This is the case, for instance, if we have $lim_{n \to \infty} \alpha'(n) = \lim_{k \to \infty} \beta'(k)$, but 
$\alpha'(n) - \beta'(n) < O(\frac{1}{n^2})$. In that case, the integral $\int_0^n \alpha'(x) - \beta'(x) \; \mathrm{d}x < k + O(\frac{1}{n})$ remains finite for $n \to \infty$ and represents the maximum value recoverable in this ecosystem which is insufficient if $k < I^C$.
}
\end{quotation}

\subsubsection{Solution of the Hub-Configuration using 2-Actor Bargaining} \label{sec-sub-hub-config-2-actor}

Based on the strategy outlined in section \ref{seq-analytic-sol-2-actors} for the 2-actor case, we will now analytically solve the $n+1$ master equations (\ref{eqn-provider-i}) and (\ref{eqn-consumer-i}) with non-vanishing $V^P_i$ under the following assumptions:
\begin{enumerate}
\item The $n$ providers $P_i$ associate an identical value $V^P_i \equiv V^P$ to their (individual) transactions with the hub consumer $C$.
\item The $n$ providers $P_i$ have the same transactions costs, $T^P_i \equiv T^P$.
\item The consumer $C$ (= hub actor) associates an identical value $V^C_i \equiv V^C$ to its individual transactions with consumer $P_i$.
\item The consumer $C$ has identical transaction costs $T^C_i \equiv T^C$ with each of the $n$ providers.
\item The consumer $C$ has initial investment costs $I^C$ prior to any transaction taking place.
\end{enumerate}
In this set-up, the $n$ individual transaction fees $X_i$ are the only economic variables which we need to consider and solve for. The preceding conditions allow us to write the master equations as follows:
\begin{eqnarray}
	V^P + X_i - T^P & > & 0, \qquad i = 1,...,n									\label{eq-hub-sol-prov-cond}	\\
	\sum_{i=1}^n V^C - \sum_{i=1}^n X_i - \sum_{i=1}^n T^C - I^C &  > & 0. 		
\end{eqnarray}
The last equation can be trivially simplified to
\begin{equation*}
	n V^C - \sum_{i=1}^n X_i - n T^C - I^C > 0.
\end{equation*}
As rational ecosystem actors, the $n$ providers will recognize that in a first best scenario their individual transaction fees need to be identical as well, i.e., $X_i \equiv X$, further simplifying the consumer feasibility condition to
\begin{equation*}
	V^C - X - T^C - \tfrac{1}{n}I^C > 0.				
\end{equation*}
This not only has reduced the master equations from $n+1$ to 2, but also allows us to rewrite them in the form of equ.~(\ref{eq-sol-space-for-x}):
\begin{equation*}
	T^P - V^P < X < V^C - T^C - \tfrac{1}{n} I^C.
\end{equation*}
These two inequalities are easily solved by using proposition \ref{p-eco-sol-2-actors} and we arrive at the first best solution for the transaction fee $X^\star$ for the hub configuration at
\begin{equation}												\label{eq-eco-sol-hub-x}
X^\star = \frac{1}{2} \left[ (V^C - V^P) - (T^C - T^P + \tfrac{1}{n} I^C) \right].		
\end{equation}

\subsubsection{General solution of the Homogeneous Hub-Configuration}

Let us now consider the case of a homogeneous hub configuration of $n$ equal providers sharing their service (or data) with a single (data or service) consumer. Contrary to the treatment in the sections above, we now no longer regard the number $n$ of participating providers as exogenous, but to be a function of the transaction fee, that is, $n = n(X)$. We also relinquish the notions that consumer benefit or consumer transaction costs scale linearly with the number of providers (e.g., $V^C = n \times \bar{V}^C$) but allow these functions to arbitrarily depend on $n(X)$, and hence, on the transaction fee: $V^C = V^C(n) = V^C \big( n(X) \big)$ and $T^C = T^C (n) = T^C(n(X))$. 

The $n+1$ master equations then read:
\begin{eqnarray}
	W^P_i = V^P + X - T^P & > & 0, \;\; i = 1,...,n(X)						  		   \label{eq-hub-prov-equal-n}	\\
	W^C = V^C \big( n(X) \big) - n(X) \cdot X - T^C \big( n(X) \big) &  > & 0. \label{eq-hub-cons-equal-n}
\end{eqnarray}

Let the aggregate (data) provider surplus be $v(X))$, where $v(\cdot)$ satisfies the envelope condition $v'(X) \equiv n(X)$. Then total welfare $W$, measured by the unweighted sum of (data) consumer benefit and (data) provider surplus, is

\begin{equation}
	W = W^C \big((n(X)\big) + v(X).
\end{equation}

The welfare-maximizing transaction fee $X^\star_W$ is then determined by $\frac{\mathrm{d}W}{\mathrm{d}X} = 0$ and, hence, fulfills the following condition:

\begin{equation}
	X^\star_W = \frac{\mathrm{d}}{\mathrm{d}n} \Big( V^C(n) - T^C(n) \Big) \Big|_{n=n(X^\star_W)}.
\end{equation}

As expected, the optimal transaction fee equals the marginal utility for the (data) consumer of adding another (data) provider.

Alternatively, we can also determine the transaction fee for a consumer maximizing its utility $W^C = W^C(n(X))$ as given by (\ref{eq-hub-cons-equal-n}). Abbreviating $\frac{\mathrm{d}n}{\mathrm{d}X} = n'(X)$, the utility-maximizing transaction fee $X^\star_C$ then satisfies
\begin{equation}				\label{eq-eco-sol-hub-x-gen}
	X^\star_C = \frac{\mathrm{d}}{\mathrm{d}n} \Big( V^C(n) - T^C(n) \Big) \Big|_{n=n(X^\star_C)} \;
	                     - \; \frac{n(X^\star_C)}{n'(X^\star_C)}.
\end{equation}
Thus, the utility-maximizing transaction fee for a transaction of a provider $P_i$ with the (single ``quasi-monopolistic'') consumer $C$ equals the consumer's marginal utility of consuming another service (or data asset) adjusted downwards by a factor related to the supply (provider) elasticity. Setting
\begin{equation*}
	\eta_P = \frac{X n'(X)}{n(X)}
\end{equation*}
we can rewrite our result in the familiar form of Lerner indices and elasticities (albeit with a sign change because we are dealing with supply-side (provider) elasticities and not the usual demand side):
\begin{equation}				\label{eq-Lerner-hub}
	\frac{X - \frac{\mathrm{d}}{\mathrm{d}n}(V^C - T^C )}{X} = - \frac{1}{\eta_P}.
\end{equation}
We easily recover the solution derived in section \ref{sec-sub-hub-config-2-actor}, eqn.~(\ref{eq-eco-sol-hub-x}) by linearizing $V^C(n)$ and $T^C(n)$ in the following form:
\begin{eqnarray*}
	V^C(n) 		& = & n \cdot \bar{V}^C,		\\
	T^C(n) 		& = & n \cdot \bar{T}^C,		\\
	n(X^\star)	& = & n^\star,					\\
	n'(X^\star) & = & \frac{2}{\Sigma W} \cdot n^\star,
\end{eqnarray*}
with $\Sigma W = (\bar{V}^C + \bar{V}^P) - (\bar{T}^C + \bar{T}^P) = W^C + W^P > 0$ being the net total value of the transaction.
\begin{equation}
	X^\star \, = \; \bar{V}^C - \bar{T}^C - \frac{1}{2} \Sigma W \; = \;
	      \dfrac{1}{2} \Big[ (\bar{V}^C - \bar{V}^P) - ( \bar{T}^C - \bar{T}^P ) \Big].
\end{equation} 

\subsection{Ecosystem Feasibility - Generic Case} \label{sec-eco-feasibility-generic-case}

In this section, we generalize the hub-based situation analysed in section \ref{subsec-eco-via-hub} to an (arbitrary) ecosystem consisting of $m$ providers $P_i$ interacting with $n$ consumers $C_j$. Let $m_i$ denote the number of consumers provider $P_i$ is interacting with, $n_j$ the number of providers consumer $C_j$ is interacting with, and $X_{ij}$ the money (transaction fee) exchanged during this transaction. Assume further that providers accrue no (fringe) benefits $V_{ij}^P$ in addition to the exchange fee $X_{ij}$, and that consumers have to invest an amount $I^C_i$ prior to the exchange.\footnote{As in section \ref{subsec-eco-via-hub}, terms in the master equations reflect aggregate values for all transactions over a certain period of time common to all participants, and not a single transaction.} 
The $\sum_{i=1}^m m_i + \sum_{j=1}^n n_k$ master equations then read 

\begin{equation}
\begin{array}{rrl}						\label{equ-master-general}
	P_i : & X_{ij} - T_{ij}^P 			  > 0, & 		\qquad j = 1,...,m_i,  \quad i=1,...,m, \\[5pt]
	C_k : & V_{kl}^C - X_{kl} - T_{kl}^C - I^C_k > 0, &	\qquad l = 1,...,n_k   \quad k=1,...,n.
\end{array}
\end{equation}

Summing up over all producer and consumer interactions we get

\begin{equation}			\label{equ-gen-eco-feasibility}
		\sum_{k=1}^n \sum_{l=1}^{n_k} \left( V_{kl}^C - T_{kl}^C \right) - \sum_{k=1}^n I_k^C -  \sum_{i=1}^m \sum_{j=1}^{m_i} T_{ij}^P > 0.
\end{equation}

Collecting the values of the inner sums into suitable ("hatted") variables then yields

\begin{equation}
	\sum_{k=1}^n \left( \hat{V}^C_k - \hat{T}^C_k - I_k^C \right) - \sum_{i=1}^m \hat{T}_i^P > 0.
\end{equation}
Replacing the remaining sums with averages over the number of consumers or providers, we can easily formulate this condition in the form of the following general ecosystem feasibility proposition.

\begin{proposition}[General ecosystem feasibility]	\label{th-gen-eco-feasibility}
An ecosystem of $n$ (data or service) consumers with average gross internal transaction benefits $\bar{V}^C$, average transaction costs $\bar{T}^C$, and initial investments of $I^C$, and $m$ (data or service) producers with average transaction costs $\bar{T}^P$ and no (fringe) benefits from transacting with consumers outside any explicit (monetary) transaction fee, is economically feasible if the following inequality holds:
\begin{equation}			
	n \cdot (\bar{V}^C - \bar{T}^C - \bar{I}^C ) - m \cdot \bar{T}^P > 0.	\label{eqn-general-sit-via}
\end{equation}
\end{proposition}

\begin{figure}[!tbp]
\begin{center}
\includegraphics[width=10cm]{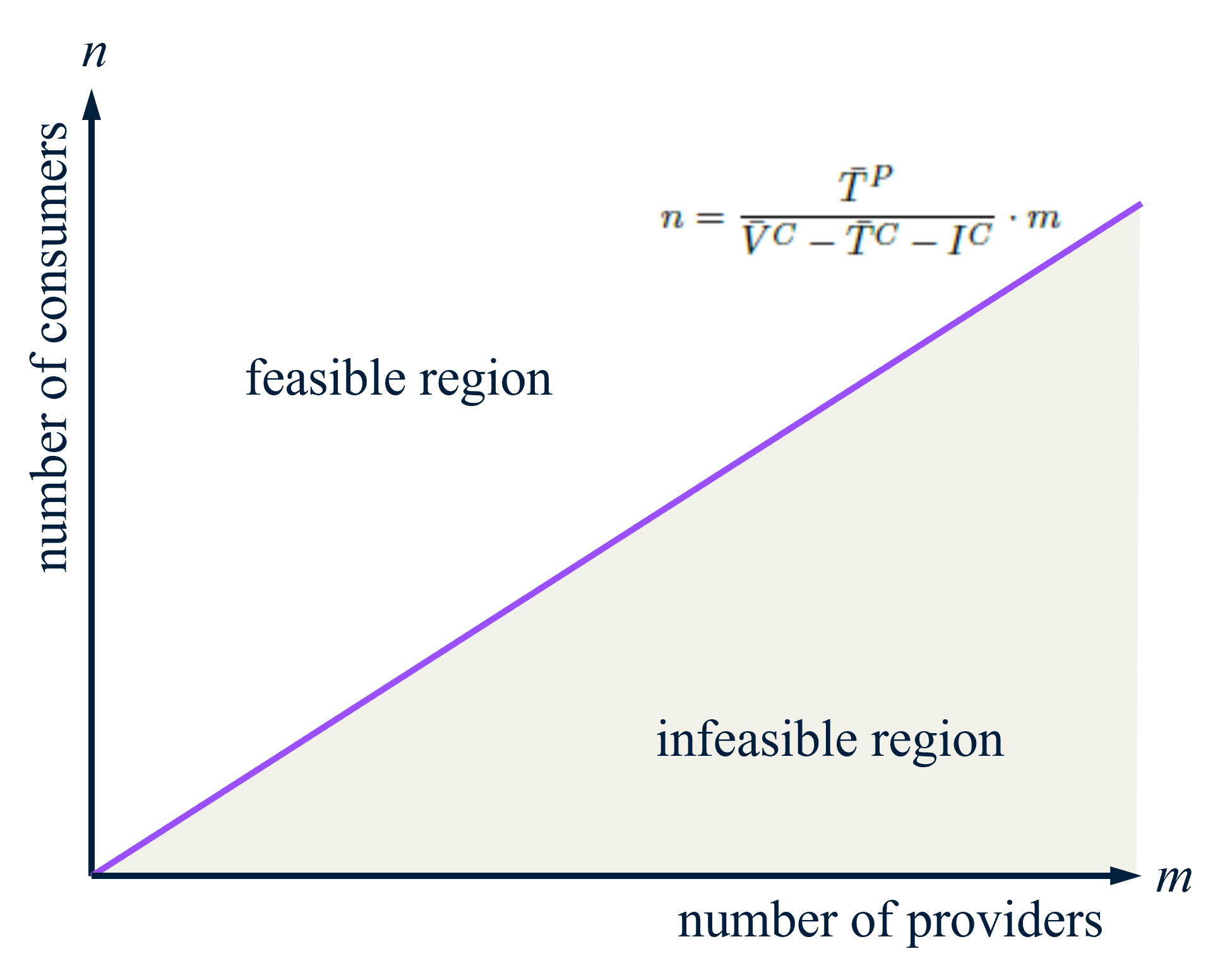}
\end{center}
\caption{Ecosystem viability diagram (generic case)}
\label{fig-general-situation-via}
\end{figure}

Equ.~(\ref{eqn-general-sit-via}) separates the $(n,m)$-plane into two regions as shown in figure \ref{fig-general-situation-via}. It is evident that increasing (average) producer transaction costs, $\bar{T}^P$, and decreasing (average) consumer value $\bar{V}^C$ of exchanges reduce the region within which the ecosystem is economically feasible. It is also clear that increasing (average) consumer transaction costs $\bar{T}^C$ and initial (\emph{ex ante}) consumer investments expand the infeasible region.

\subsection{Ecosystem Feasibility - General Formulation}

Evidently, our analytical expressions such as eqn.s~(\ref{eqn-eco-platform}) or (\ref{equ-gen-eco-feasibility}) get somewhat unwieldy for larger ecosystems. We overcome this by an algebraic model of \emph{ecosystems} using the well-established concept of an adjacency matrix for (interaction) networks from complexity theory (see, e.g., \cite{Thurner-2018}) which also has been successfully used in modelling ecological networks in biology \cite{Landi-2018}.

\begin{definition}[Ecosystem]
An (economic or business\footnote{Our analysis does not extend to biological ecosystems or ecological networks because the entries in our matrix $R$ are typically incompatible with their biological counterparts. See, \cite{Landi-2018} for a review.} \emph{\textbf{ecosystem}} $\mathscr{E}$ is a structure $\mE = \langle M, R, V, X, T \rangle_t$ with
\begin{enumerate}
\item $M,$ the set of ecosystem participants,\footnote{Even if we allow the case of an empty ecosystem in this definition, i.e., $M = \emptyset$, we are not really interested in such pathological situations in this paper.}
\item $R \subseteq M \times M$, the set of ecosystem relations. If $(p,c) \in R$, we call $p$ a \emph{provider} and $c$ a \emph{consumer},
\item $V : R \to  \mathbb{R}^+_0 \times \mathbb{R}^+_0$, a function indicating the provider and consumer benefits for a particular transaction (with $\mathbb{R}^+_0$ denoting the positive reals including $0$). We will sometimes write $V((a,b)) = (V_{ab}^P, V_{ab}^C)$  to explicitly indicate provider and consumer roles.
\item $X : R \to \mathbb{R}$, a function indicating the transaction fee $X((a, b))$ for a transaction between participants $a$ and $b$.
\item $T : R \to  \mathbb{R}^+_0 \times \mathbb{R}^+_0$, a function indicating transaction costs $T((a,b)) = (T_{ab}^P, T_{ab}^C)$ for the respective transaction between participants $a$ and $b$.
\end{enumerate}
The subscript $t$ in $\langle M, R, V, X, T \rangle_t$ indicates that, strictly speaking, this particular form of describing an ecosystem is only valid for a single point in time (\emph{viz}., for a single transaction) or for a specific time interval. In the latter case, values $V$, $T$, and $X$ then represent the sum of all transactions conducted within this time period.

For convenience, we also define the set $P$ of all providers, and the set $C$ of all consumers as $P = \{ p \, | \,  \exists c \in M: (p,c) \in R \}$ and 
$C = \{ c \, | \, \exists p \in M: (p,c) \in R \}$.
\end{definition}

We will also make use of the following $\Sigma$-operator:

\begin{definition}[Sigma operator] \label{def-sigma-operator}
Let $A$ be a (finite\footnote{We only need finiteness of $A$ as a lazy way to guarantee convergence.}) set and $f:A \to \mathbb{R}$ a (valuation) function from $A$ to the set of reals. Then we define the \emph{\textbf{Sigma operator}} $\Sigma[\cdot, \cdot]$ as follows.
\begin{equation}				\label{eq-def-sigma}
 \Sigma[ A, f] = \sum_{a \in A} f(a) \; \in \mathbb{R}.
\end{equation}
\end{definition}
Later we will make use of the following corollary.
\begin{corollary}		\label{col-linear-sigma}
Let $A$ be a (finite) set, $f: A \to \mathbb{R}$ and $g: A \to \mathbb{R}$ arbitrary (valuation) functions, and $\Sigma$ the Sigma-Operator as defined in  definition \ref{def-sigma-operator} above. Then $\Sigma$ is linear in its second argument as follows:
\begin{equation}
	\Sigma [A, f + g] = \Sigma [A, f] + \Sigma [A, g].
\end{equation}
\end{corollary}
\begin{proof}
We have 
\begin{align*}
	\Sigma [A, f + g] = \sum_{a \in A} (f+g)(a) & = \sum_{a \in A} \left( f(a) + g(a) \right) 					\\
								& = \sum_{a \in A} f(a) + \sum_{a \in A} g(a) 
								  = \Sigma [A, f] + \Sigma [A, g].
\end{align*}
\end{proof}

We also profit from some auxiliary definitions.

\begin{definition}[Projectors and projections]
Let $\mE = \langle M, R, V, X, T \rangle_t$ be an ecosystem. Then we define the natural projection functions $\pi^P$ and $\pi^C$ to select \emph{producers} and \emph{consumers} as follows:
\begin{align*}
	\pi^P : \mathbb{R} \times \mathbb{R} 	& \to \mathbb{R}			\\
	                          (a, b)		& \mapsto a		\\[4pt]
	\pi^C : \mathbb{R} \times \mathbb{R}    & \to \mathbb{R}			\\
                    	        (a, b)      & \mapsto b	
\end{align*}
We then abbreviate:
\begin{align*}
	V^P : R \to \mathbb{R}^+_0,	& \quad V^P = \pi^P \circ V,		\\
	V^C : R \to \mathbb{R}^+_0,	& \quad V^C = \pi^C \circ V,		\\
	T^P : R \to \mathbb{R}^+_0,	& \quad T^P = \pi^P \circ T,		\\
	T^C : R \to \mathbb{R}^+_0,	& \quad T^C = \pi^C \circ T.
\end{align*}
\end{definition}

Using this apparatus we can easily rewrite the \textbf{master equations} for the general ecosystem case as follows:

\begin{proposition}[Master equation - internal feasibility]
Let $\mE$ be an ecosystem, i.e., $\mE = \langle M, R, V, X, T \rangle_t$. Then $\mE$ is \emph{internally feasible} if the sum of net benefits (i.e., $V-T$) of providers and consumers is positive.
\begin{equation}
\Sigma \left[R, V^P - T^P \right] + \Sigma \left[ R, V^C - T^C \right] > 0.
\end{equation}
\end{proposition}
\begin{proof}
This proposition is the slight generalization of the derivation of the master equations (\ref{equ-master-general}) and (\ref{equ-gen-eco-feasibility}) with non-vanishing $V_{ij}^P$ as given in section \ref{sec-eco-feasibility-generic-case}.

Let $(a,b) \in R$ be an arbitrary provider-consumer interaction of $\mE$. Then $V((a,b)) = (V^P_{ab}, V^C_{ab})$ and $T((a,b)) = (T^P_{ab},T^C_{ab})$ are the expected values and transaction costs for provider $a$ and consumer $b$, respectively. Denoting the transaction fee of this interaction by $X_{ab}$, the two \emph{master equations} for this transaction are given by
\begin{align}
	V^P_{ab} + X_{ab} - T^P_{ab} & \; > \;  0, 	\label{eq-master-prov-ab}		\\
	V^C_{ab} - X_{ab} - T^C_{ab} & \; > \;  0. 	\label{eq-master-cons-ab}	
\end{align}
Adding (\ref{eq-master-prov-ab}) and (\ref{eq-master-cons-ab}) we get
\begin{equation}
	V^P_{ab} - T^P_{ab} + 	V^C_{ab} - T^C_{ab} \; > \; 0.			\label{eq-master-prov-cons-ab}
\end{equation}
We now have to sum equ.~(\ref{eq-master-prov-cons-ab}) over all existing transaction pairs $(a,b) \in R$ to obtain the \emph{master equation} for the whole ecosystem $\mE$:
\begin{equation}							\label{eq-master-prov-cons}
	\sum_{(a,b) \in R} \left( V^P_{ab} - T^P_{ab} \right) \; +  \sum_{(c,d) \in R} \left( V^C_{cd} - T^C_{cd} \right) \; > \; 0.
\end{equation}
Observing
\begin{align*}
	V^P_{ab} = \pi^P(V((a,b)) = \pi^P \circ V ((a,b)) = V^P((a,b),	\\
	V^C_{ab} = \pi^C(V((a,b)) = \pi^C \circ V ((a,b)) = V^C((a,b),
\end{align*}
for the expected values $V_{ab}$ for a provider and consumer, and similarly, for transaction costs $T$, we can write (\ref{eq-master-prov-cons}) as
\begin{align*}
	\sum_{(a,b) \in R}  V^P((a,b))  - 	\sum_{(a,b) \in R}  T^P((a,b))  & 	\\
	+ \sum_{(a,b) \in R}  V^C((a,b))  - 	\sum_{(a,b) \in R} T^C((a,b))  & \; > \; 0.
\end{align*}
We rewrite this using our $\Sigma[\cdot,\cdot]$ operator
\begin{equation*}
	\Sigma[ R, V^P] - \Sigma[ R, T^P] + \Sigma[ R, V^C] - \Sigma[ R, T^C] \; > \; 0,
\end{equation*}
and use the linearity of $\Sigma$ to further simplify the expression to get the required
\begin{equation*}
	\Sigma \left[R, V^P - T^P \right] + \Sigma \left[ R, V^C - T^C) \right] > 0.
\end{equation*}
\end{proof}

Finally, we will use proposition \ref{p-eco-sol-2-actors} to give an analytical solution to the ecosystem master equations as given by (\ref{eq-master-prov-ab}) and (\ref{eq-master-prov-cons-ab}).

\begin{proposition}[Generic ecosystem solution]
Let $\mE = \langle M, R, V, X, T \rangle_t$ be an ecosystem consisting of rational ecosystem actors. Then the solutions to the master equations for transactions $(a,b) \in R$ in a first-best setting are given by
\begin{equation}
	X^\star_{ab} = \frac{1}{2} \left[ (V^C_{ab} - V^P_{ab}) - (T^C_{ab} - T^P_{ab}) \right].
\end{equation}
\end{proposition}
\begin{proof}
The only thing we have to show is that the set of individual 2-actor solutions, $X^\star_{ab}$, is indeed the first-best solution for solving all master equations together. We will do this in an indirect manner.

Assume that for some transaction $(r,t)\in R$ the transaction fee deviates from our solution, that is, $X_{rt} \neq X^\star_{rt}$. Take the case where $X_{rt} > X^\star_{rt}$. Here, provider $r$ extracts more value $W^P_{rs}$ than in our $\star$-solution. Because of the first-best setting, another provider $s$ would realize and herself try to deviate in her favour from the transaction fee with consumer $t$, leading to further ``defectors'' for transactions with consumer $t$. Noticing that such a deviation is possible for one consumer, already defected providers will now try to do the same with other consumers, further spreading the defection. Evidently, such a strategy is not stable under these conditions and, hence, providers will not follow it.

The same argument holds in the other case where $X_{rs} < X^\star_{rs}$, that is, where consumer $s$ is able to extract more value from the transaction than in our $\star$-solution.

Therefore, any deviation of $X_{rs}$ from our equilibrium solution $X^\star$ is unstable and will not be implemented by our rational economic actors in a first best setting.
\end{proof}

\section{Discussion}		\label{sec-discussion}

\subsection{Value co-creation}

The defining feature of our approach to (business) ecosystem modeling of an ecosystem transaction between actor $i$ and actor $j$ is the inclusion of an additional source of value for the ``provider'' $i$, $V^P_{ij}$:
\begin{eqnarray*}
	\mathrm{Provider: ~~Actor~} i \; & W^P_{ij} = V^P_{ij} + X_{ij} - T^P_{ij} & > \ 0, 			\\
	\mathrm{Consumer: ~~Actor~} j \; & W^C_{ij} = V^C_{ij} - X_{ij} - T^C_{ij} & > \ 0.	
\end{eqnarray*}
This benefit accrues in excess of the immediate value of the transaction (namely, the transaction fee $X_{ij}$) for the provider. Such a situation occurs, for instance, in  manufacturing ecosystems, where many different users of a particular machine (e.g., a milling machine) generate usage data which the machine manufacturer may collectively use for applying federated learning techniques in order to (later) provide predictive maintenance services to the original data providers. Additionally, participants in a certain value chain may want to share some data such as service descriptions or production schedules with other participants in order to increase the overall calamity resilience of the network \cite{Dumss-2021, Weber-2022}.

The presence of $V^C$ on the consumer side is less debatable as this term simply captures the underlying business rationale for conducting the transaction at all. Because many TCE-scenarios are more focused on de\-ter\-mi\-ning the optimal governance structure by comparing transaction costs alone, these terms are not needed for the analysis and, hence, are not present.

The (superficial) antisymmetry between ``providers'' and ``consumers'' in the otherwise symmetrical master equations can be easily explained (away) by the money flow, i.e., the transaction exchange fee $X$. In our nomenclature, the party of a (2-actor) ecosystem transaction which pays out or disburses a (positive) transaction fee is called the ``consumer'', rendering the other party automatically a ``provider''. For those cases where our solution also allows negative transaction fees, $X^\star < 0$, we can either simply exchange labels $C$ and $P$ to restore symmetry or regard the ``transaction fee'' a subsidy.

\subsection{Implications of the first-best transaction fee}

Let's now discuss the implications of the equilibrium transaction fee, $X^\star$, in the first-best case of rational ecosystem actors. For simplification reasons and because of its generality, we drop the subscripts $i$ and $j$ of provider and consumer.

\begin{equation*}												
X^\star = \frac{1}{2} \left[ (V^C - V^P) - (T^C - T^P) \right].		
\end{equation*}
A quick economic analysis of this formula yields the following observations:
\begin{itemize}
\item There will not be any transaction fee in a completely egalitarian ecosystem with $V^P \equiv V^C$ and $T^P \equiv T^C$.
\item Rewriting the equation above as
\begin{equation*}
X^\star = \frac{1}{2} \left[ (V^C - T^C) - (V^P - T^P) \right]	,	
\end{equation*}
one easily recognizes that the size of the transaction fee is determined by the difference between the net benefits (net values) of the transaction for the consumer, $V^C - T^C$, and the provider, $V^P - T^P$.

\item In hub-and-spoke ecosystems with a single consumer $C$ interacting with a lot of providers $P$, the consumer will have to allocate a significant part (50\% in our first-best setting) of its overall value of the ecosystem (more or less evenly) to the individual providers (at the end of the ``spokes'').
\item The lower the transaction costs for the consumer, $T^C$, the higher the trans\-action fee. This supports one reason, why (IT-)platform-based ecosystems are so attractive for providers.
\item If provider and consumer transaction costs are identical, $T^C \equiv T^P$, as, for instance, is the case in data spaces \cite{Strnadl-2023b}, they no longer play a role in determining the transaction fee. $X^\star$ is then only determined by the gross value difference $\Delta V \equiv V^C - V^P$. We need not forget, though, that total transaction costs need to be recoverable by the total value of the transaction as per eqn.\,(\ref{eqn-tx-feasibility-general}).
\item In case the provider does not expect any additional value from the transaction other than the transaction fee, $V^P = 0$ (that is, they do not work in an ecosystem mode), the consumer value $V^C$ has to finance a part (in the first-best case: 50\%) of provider transaction costs.
\item The formula in principle allows negative transaction fees. This apparent asymmetry can be easily eliminated by either simply exchanging the provider and consumer labels $P$ and $C$ restoring the full symmetry again or regarding the transaction fee a subsidy.
\end{itemize}

\subsection{Economic value of an ecosystem}

Simplifying our notation, the master equation for ecosystem feasibility may be denoted as (recall that $P$ and $C$ are the sets of providers and consumers, respectively):
\begin{equation*}
	\sum_{i \in P} W^P_i + \sum_{j \in C } W^C_j \; > \; 0.
\end{equation*}
This translates to the (necessary) condition that the total economic value of the ecosystem produced by providers and consumers needs to be positive in order for the ecosystem to be stable. There are two potential sources of value
\begin{enumerate}
\item from within the ecosystem: This typically arises in form of cost savings for participants.
\item from outside the ecosystem: This may occur in form of additional revenues for (some) participants obtained from enterprises outside the ecosystem.
\end{enumerate}
This interpretation aligns very well with Adner's definition of an ecosystem \cite{Adner-2017} highlighting the need for a ``focal value proposition'' to materialize. If such a value source cannot be identified, ecosystems will not emerge and the intended economic activity will be conducted within other forms of governance (\emph{viz.}, a firm or via the market mechanism).

\subsection{Role of an ecosystem federator}

Up to now our model of an ecosystem only included participants in the form of (data or service) providers and (data or service) consumers. As is well known \cite{Strnadl-2023a}, many if not all ecosystems require some form of coordination in the form of a dedicated entity which we will call \emph{federator}. The federator then provides common ecosystem infrastructure services such as on-boarding, identity and access management services, catalog services, and, depending on the particular ecosystem and industry, others as well. 

Federators recoup the costs of providing these services to ecosystem participants by charging the participants various forms of fees such as (on-time) onboarding fees, recurring (fixed) subscription fees, or (fixed or variable) transaction fees. Let's assume, for simplicity reasons, an ecosystem where the federator charges both, providers and consumers, a fixed fee, $F^P$ and $F^C$, which might differ for the two types of participants, for every transaction they conduct in order to recover transaction costs (which may include earning back initial investments) $T^C$. 
Then the master equations need to be extended as follows:
\begin{eqnarray}
	\mathrm{Provider:} \; & W^P = V^P + X - T^P - F^P & > \ 0, 			\\
	\mathrm{Consumer:} \; & W^C = V^C - X - T^C - F^C & > \ 0,			\\
	\mathrm{Federator:} \; & W^F = F^P + F^C - T^F > 0.				\label{eq-federator-feasibility}
\end{eqnarray}
It is easy to see that in this model all generic results of our analysis are still valid when we (i) simply use slightly modified transaction costs $T' = T + F$ for both, providers and consumers (hence, we have dropped the superscripts), and (ii) observe the ancilliary (participation) condition $W^F > 0$ for the federator as per 
eqn.\,(\ref{eq-federator-feasibility}).

\subsection{Transaction fee comparison of Gaia-X-based ecosystems and data spaces}

In Gaia-X-based ecosystems, service providers have to incur higher transaction costs because of the need to identify and authenticate service consumers. Because currently relevant Gaia-X standards are predominantly incompatible with existing authentication and authorization (A\&A) methods of corporate IT systems, service providers have to install and operate complex internal A\&A translation services in order to bridge the Gaia-X world with the corporate world. This is in stark contrast to classical data spaces concepts like the IDS RAM 3.0 where the set up of (data) providers and (data consumers) is completely symmetrical and handled by a dedicated (IT) component, called a ``data space connector''.

Let us model this situation by setting Gaia-X consumer costs $T^C_g$ at a fraction $\alpha$ of provider costs, 
$T^C_g = \alpha T^P_g$, with $0 < \alpha < 1$ typically in the range of 0.01 to 0.1, and set data space transaction costs $T^P_d = T^C_d \equiv T_d$, and  subscripts $g$ and $d$ denoting Gaia-X or data space contexts, respectively. Assuming transaction values are identical for provider and consumer irrespective of whether the transaction is conducted in a Gaia-X ecosystem or a data space, the equitable bargaining solutions $X^\star_g$ and $X^\star_d$ are given by
\begin{eqnarray}
	X^\star_g & = & \frac{1}{2} \big[ \Delta V + (1 - \alpha)T^P_g \big],		\\
	X^\star_d & = & \frac{1}{2} \Delta V.
\end{eqnarray}
That essentially means that equitably negotiated transaction fees in a Gaia-X ecosystem are higher than in a corresponding data space. This is due to the fact that the bargaining solution allows the higher provider costs to be covered by the transaction fee. We also note the following comments here:
\begin{itemize}
\item The comparison of a pure data space and Gaia-X-based ecosystems is inasmuch inappropriate as many current data space standards are limited to the exchange or transmission of data assets whereas Gaia-X ecosystems allow the provision of arbitrary (IT) services ranging from IaaS, PaaS to any full-fledged SaaS solution.
\item The derivation of $X^\star$ assumes equal bargaining power for both, pro\-vi\-der and consumer. We encounter such a situation in B2B data spaces, but it is not the prevalent way how XaaS services are generally offered where XaaS ``platforms'' typically deliver the respective services. In such a decidedly more hub-and-spoke configuration, the assumption underlying the applicability of our $X^\star_g$ may not apply to the same extent as in the case of data spaces.
\item On a side note, installation and operation of the so-called \emph{data space connectors} mentioned above is, in general, more complex and cost-intensive than the IT-components providing Gaia-X identities. This massively affects (data) provider costs, $T^P$, as we have $T^P_d = \beta T^P_g$ with 
$\beta$ in the range 30-60\%, which is larger than $\alpha$ by 1 or 2 orders of magnitude.
\end{itemize}

\subsection{Structural indeterminacy of ecosystems}

By construction, our model of an ecosystem, $\mE = \langle M, R, V, X, T \rangle_t$,
has the structure of a directed graph by virtue of the relationship matrix $R \subset M \times M$ with $M$ the set of nodes and $R$ the set of vertices or edges of the graph. If we disregard (for the moment) the ecosystem's internal value function $V$ --- that is, we just consider a structure $\langle M, R, X, T \rangle$ --- one clearly cannot discriminate the resulting construct from a (more or less integrated) value chain or any other (contingent or not) market arrangement. We interpret this result in form of the following (informal) proposition:

\begin{proposition}[Structural indeterminacy of ecosystems]
Ecosystems cannot be identified by structural properties (i.e., the interaction matrix of ecosystem participants) alone.
\end{proposition}

This is consistent with and corroborates Jacobides' definition of ecosystems \cite{Jacobides-2018} who puts forward the existence of non-generic complementarities of ecosystem actors as a defining feature. Consequently, two-sided platforms are not regarded as an ecosystem \emph{proper} because platform actors typically do not share specific complementarities at all. 

We can easily map this thinking onto our framework where the structure (the graph) of a two-sided platform, irrespective of whether an ecosystem in the sense of Jacobides or not, always has a star-shaped hub-and-spoke to\-po\-lo\-gy with the platform provider as a hub. Answering the question whether this is an ecosystem or not then reduces to investigating whether platform participants share a common \emph{value function} $V$ or not. Only in case of a shared value proposition (captured in our function $V$), we may identify the platform also as an ecosystem \emph{proper}.

\subsection{Future research}
We can include the \emph{platform operator} or \emph{Gaia-X federator} into our economic analysis. At this stage of the formalization, any federator fees are absorbed by or included in the transaction costs, $T$, and, hence, not available for direct analysis.

Additionally, one may expand the current work by introducing personal utility scales and re-phrase the master equation in terms of utility maximization of individual ecosystem participants.

\subsection*{Acknowledgements}

I am highly appreciative of the repeated and deep discussions with Lucas Eustache and Surjasama Lahiri of Universit\' e Paris Dauphine-PSL. We are currently fleshing out the theory in more details in order to submit it to a suitable academic journal.

We also gratefully acknowledge early discussions on select parts of the initial version of this work with Assoc.~Prof.~Dr.~Andreas Mild of Vienna University of Economics and Business. 

We also thank a scientific journalist (SP) for a critical reading of the work albeit this was strictly limited to the English language with no bearing on the material content.

\subsection*{Declarations}

\textbf{Funding:} The presented work has been partly funded by the research project iECO (intelligent Empowerment of Construction Industry) which is supported by the German Federal Ministry for Economic Affairs and Climate Action (BMWK), grant no. 68GX21011A, and monitored by the Bundesnetzagentur, and partly by the research project EuProGigant supported by the Austrian Research Promotion Agency (grant no. ID 883413) and the German Federal Ministry for Economic Affairs and Energy (grant no. FKZ 01MJ21008A).

\textbf{Financial interests:} The author declares he has no financial interests.

\textbf{Non-financial interests:} The author is an unpaid member of the Gaia-X Architecture Working Group and an unpaid member of the Mobility Data Space Technical Committee.

%

\bibliographystyle{ieeetr}

\bibliography{Gaia-X_Value_Analysis}

\end{document}